\documentclass{iaus}
\usepackage{graphicx}
\usepackage{natbib}

\title[Terrestrial Planets Formation] %% give here short title %%
{Simulations for Terrestrial Planets Formation}

\author[Ji \& Zhang]   %% give here short author list %%
{Jianghui JI$^{1}$, \and Niu  ZHANG$^{1,2}$,}

\affiliation{$^1$Purple  Mountain  Observatory, Chinese  Academy
of  Sciences,  Nanjing  210008, China \break email: jijh@pmo.ac.cn\\[\affilskip]
$^2$Graduate School of Chinese Academy of Science, Beijing 100049 }

\pubyear{2009}
\volume{263}  %% insert here IAU Symposium No.
\pagerange{1--5}

\jname{Icy Bodies of the Solar System}

\editors{Julio A. Fernandez,Daniela Lazzaro, \& Dina
Prialnik-Kovetz}
\begin{document}

\maketitle

\begin{abstract}
We investigate the formation of terrestrial planets in the late
stage of planetary formation using two-planet model. At that time,
the protostar has formed for  about 3 Myr and the gas disk has
dissipated. In the model,  the perturbations from Jupiter and Saturn
are considered. We also consider variations of the mass of outer
planet, and the initial eccentricities and inclinations of embryos
and planetesimals. Our results show that, terrestrial planets are
formed in 50 Myr, and the accretion rate is about 60\% - 80\%. In
each simulation, 3 - 4 terrestrial planets are formed inside
"Jupiter" with masses of  $0.15 - 3.6 M_{\oplus}$. In the 0.5 - 4AU,
when the eccentricities of planetesimals are excited, planetesimals
are able to accrete material from wide radial direction. The plenty
of water material of the terrestrial planet in the Habitable Zone
may be transferred from the farther places by this mechanism.
Accretion may also happen a few times between two giant planets only
if the outer planet has a moderate mass and the small terrestrial
planet could survive at some resonances over time scale of $10^8$
yr.

\end{abstract}

\firstsection % if your document starts with a section,

\section{Introduction}
The discovery of the extrasolar planets \citep{May95,Lee02,Ji03}
around solar-type stars indeed provides substantial clues for the
formation and origin of our own solar system. According to standard
theory \citep{Saf69,Wet90,Lis93}, it is generally believed that
planet formation may experience such several stages: in the early
stage, the dust grains condense to grow km-sized planetesimals; in
the middle stage, Moon-to-Mars sized embryos are created by
accretion of planetesimals. When the embryos grow up to a core of
$\sim 10M_\oplus$, runaway accretion may take place. With more gases
accreted onto the solid core, the embryos become more massive and
eventually collapse to produce giant Jovian planets \citep{Ida04}.
At the end of the stage, it is around that the protostar has formed
for about $3$ Myr, the gas disk has dissipated. A few larger bodies
with low  $e$ and $i$ are in crowds of planetesimals with certain
eccentricities and inclinations. In the late stage, the terrestrial
embryos are excited to high eccentricity orbits by mutual
gravitational perturbation. Next, the orbital crossings make planets
obtain material in wider radial area. In this sense, solid residue
is either scattered out of the planetary system or accreted by the
massive planet, even being captured \citep{nag00} at the resonance
position of the giant planets.

\citet{cha01} made a study of terrestrial planet formation in the
late stage by numerical simulations, who set $150 - 160$
Moon-to-Mars size planetary embryos in the area of $0.3 - 2.0$ AU
under mutual interactions from Jupiter and Saturn. He also examined
two initial mass distributions: approximately uniform masses, and a
bimodal mass distribution. The results show that $2 - 4$ planets are
formed within $50$ Myr, and finally survive over $200$ Myr
timescale, and the final planets usually have eccentric orbits with
higher eccentricities and inclinations .  \citet{ray04, ray06} also
investigated the formation of terrestrial planets. In the
simulations, they simply took into account Jupiter's gravitational
perturbation, and the distribution of material are in $0.5 - 4.5$
AU. Their results confirm a leading hypothesis for the origin of
Earth's water: they may come from the material in the outer area by
impacts in the late stage of planet formation. \citet{ray06b}
explored the planet formation under planetary migration of the
giant. In the simulations, super Hot Earth form interior to the
migrating giant planet, and water-rich, Earth-size terrestrial
planet are present in the Habitable Zone ($0.8 - 1.5$ AU) and can
survive over $10^8$ yr timescale.

In our work, we consider two-planet model, in which Jupiter and
Saturn are supposed to be already formed, with two swarms of
planetesimals distributed in the region among $0.5 - 4.2$ AU and
$6.2 - 9.6$ AU respectively. The initial eccentricities and
inclinations of planetesimals are considered. We also vary the mass
of Saturn to examine how the small bodies evolve. The simulations
are performed on longer timescale $400$ Myr in order to check the
stability and the dynamical structure evolution of the system. In
the following, we briefly summarize our numerical setup and results
.

\section{Numerical Setup}
The timescale for formation of Jupiter-like planets is usually
considered to be less than $10$ Myr \citep{bri01}, the formation
scenario of planet embryos is related to their heliocentric
distances and the initial mass of the star nebular. If we adopt the
model of $1.5$ MMSN (Minimum Mass Solar Nebular), the upper bound of
the timescale for Jupiter-like planet formation corresponds to the
timescale for embryo formation at $2.5$ AU \citep{kok02}, which is
just at $3:1$ resonance location of Jupiter. In the region $2.5 -
4.2$ AU, embryos will be cleared off by strong perturbation from
Jupiter. There should be some much smaller solid residue among
Jupiter and Saturn, even though the clearing effect may throw out
most of the material in this area. We set embryos simply in the
region $0.5 - 2.5$ AU and planetesimals at $0.5 - 4.2$ AU and $6.2 -
9.6$ AU.

We adopt the surface density profile as follows \citep{ray04}:
\begin{equation}
\Sigma (r) = \left\{
\begin{array}{ll}
\Sigma_{1}r^{-3/2}, & r < snow~line,\\
\Sigma_{snow}(\frac{r}{5AU})^{-3/2}, & r > snow~line.
\end{array}
\right.
\end{equation}

In (2.1), $\Sigma_{snow} = 4~g/cm^2 $ is the surface density at
snowline, where the snowline is at $2.5$ AU with $\Sigma_1 =
10~g/cm^2 $. The mass of planetary embryos is proportional to the
width of the feeding zone, which is associated with Hill Radius,
$R_{H} $, so the mass of an embryo increases as
\begin{equation}
M_{embryo} \propto r\Sigma(r) R_{H}
\end{equation}

The embryos in the $0.5 - 2.5$ AU are spaced by $\Lambda $ ($\Lambda
$ varying randomly between 2 and 5) mutual Hill Radii, $R_{H,m}$,
which is defined as
\begin{equation}
R_{H,m} = (\frac{a_1+a_2}{2})(\frac{m_1+m_2}{3M_{\odot}})^{1/3}
\end{equation}
where $a_{1,2} $ and $m_{1,2} $ are the semi-major axes and masses
of the embryos respectively. Replacing $R_{H} $ in (2.2) with
$R_{H,m}$, and substituting (2.1) in (2.2), then, we achieve a
relation law between the mass of embryos and the parameter $\Lambda
$ as
\begin{equation}
M_{embryo} \propto r\Sigma(r) R_{H,m} \propto
r^{3/4}\Lambda^{3/2}\Sigma^{3/2}
\end{equation}

Here, we equally set the masses of planetesimals inside and outside
Jupiter, respectively. Consequently, the number distribution of the
planetesimals is simply required to meet $N\propto r^{-1/2} $.
Additionally, we remain the total number of planetesimals and
embryos inside Jupiter, and the number of planetesimals outside
Jupiter both equal to $200$. The mass inside and outside Jupiter is
equal to be $7.5M_\oplus $. The eccentricities and inclinations vary
in ($0 - 0.02$) and ($0 - 0.05^\circ$), respectively.  The mass of
Saturn in simulations 1a/1b, 2a/2b and 3a/3b  are $0.5M_\oplus $,
$5M_\oplus $, $50M_\oplus $ respectively. Each simulations marked by
label a (or b) is run to consider (not consider) self-gravitation of
planetesimals among giants.

\begin{figure}
\includegraphics[height=2.0in, width=2.4in]{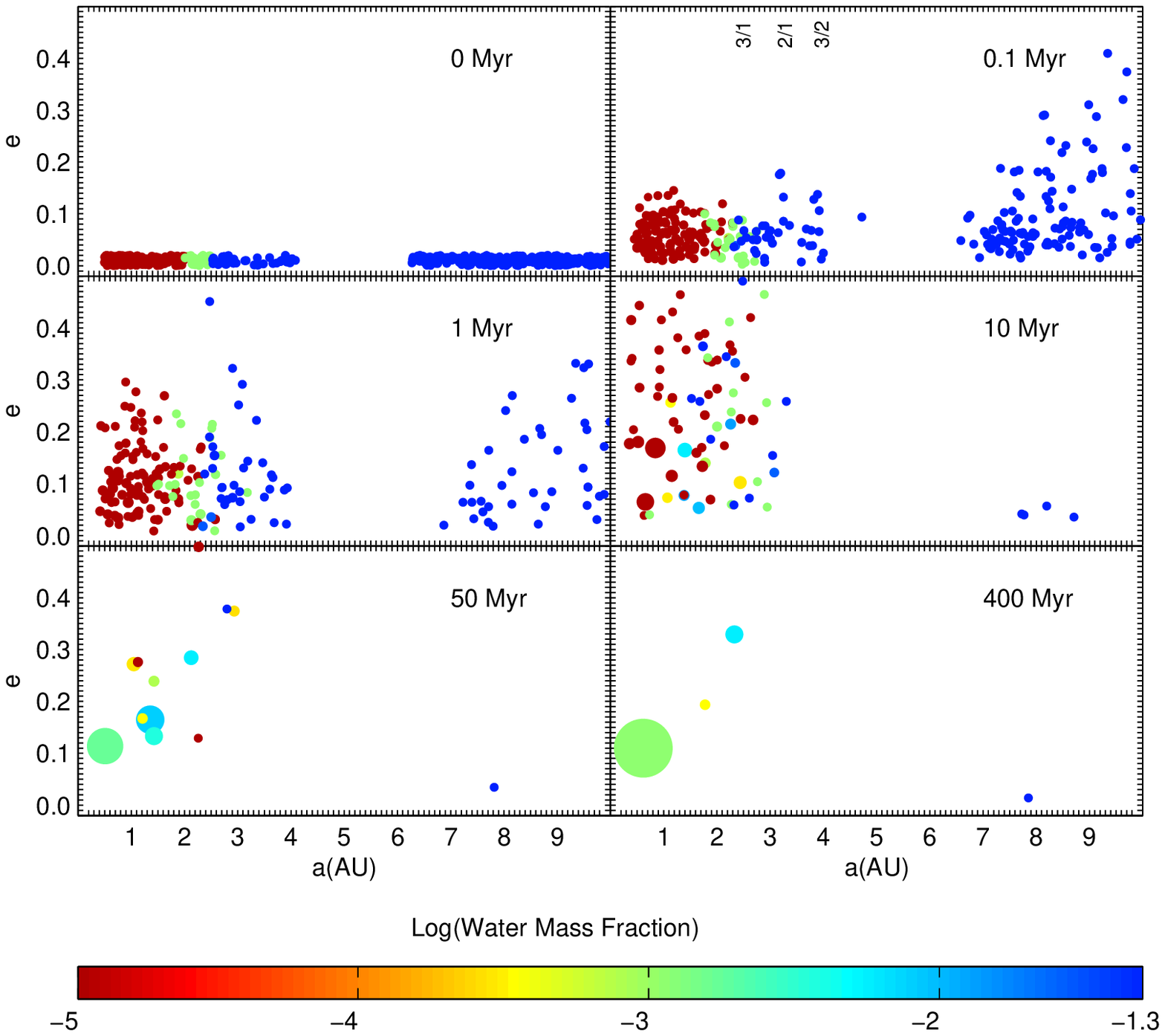}
\includegraphics[height=2.0in, width=2.4in]{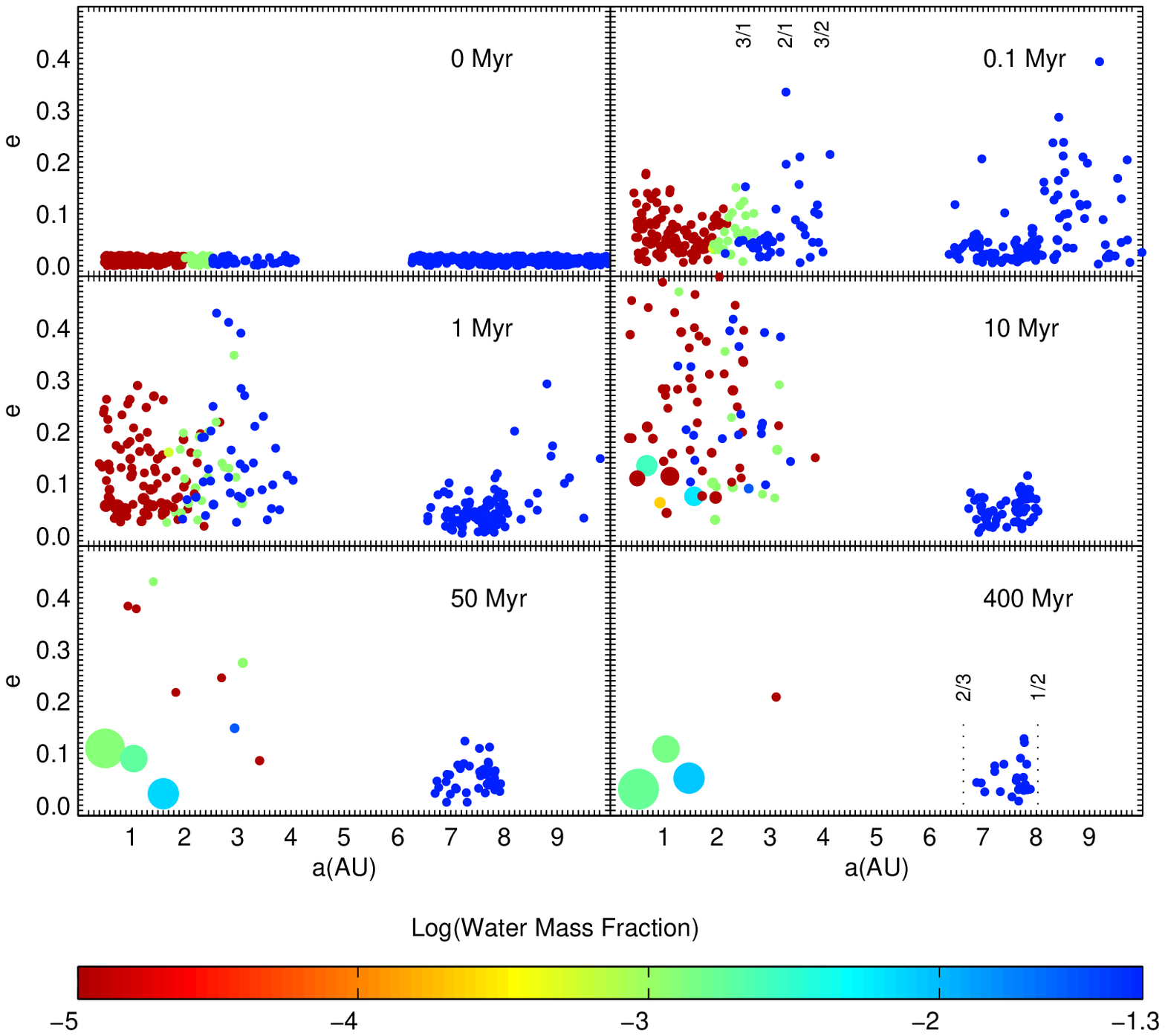}
  \caption{\textit{Left panel}: (a)
  Snapshot of simulation 2a with $M_{Saturn}=5M_{\oplus}$.
  The total mass of embryos is
$2.4M_{\oplus}$, the masses of planetesimals inside Jupiter are
$0.0317M_{\oplus}$, and those outside Jupiter are $0.0375M_{\oplus}
$. Planetesimals among Jupiter and Saturn were nonself-gravitational
(see Section 2.1). Note the size of each object is relative, and the
value bar is log of water mass fraction.
  \textit{Right panel}: (b) snapshot for similuation 2b.
} \label{fig1}
\end{figure}

We use the hybrid symplectic integrator \citep{cha99} in MERCURY
package to integrate all the simulations.  In addition, we adopt $6$
days as the length of time step, which is a twentieth period of the
innermost body at $0.5$ AU. All runs are carried out over $400$ Myr
time scale. At the end of the intergration, the changes of energy
and angular momenta are $10^{-3}$ and $10^{-11}$ respectively. Six
simulations are performed on a workstation composed of $12$ CPUs
with $1.2$ GHz, and each costs roughly $45$ days.

\section{Results}
Fig. 1(a) is a snapshot of simulation 2a. At $0.1$ Myr, it is clear
that the planetesimals are excited at the $3:2$ ($3.97$ AU),$2:1$
($3.28$ AU) and $3:1$ ($2.5$ AU) resonance positions with Jupiter,
and this is quite similar to the Kirkwood gaps of the main
asteroidal belt in solar system. For about $1$ Myr, planetesimals
and embryos are deeply intermixed, where most of the bodies have
stirred to be large eccentricities. Collisions and accretions
frequently emerge among planetesimals and embryos. This process
continues until about $50$ Myr, and the planetary embryos are mostly
generated. The formation timescale of embryos is in accordance with
that of \citep{Ida04}. Finally, inside Jupiter, $3$ terrestrial
planets are formed with masses of $0.15 - 3.6 M_{\oplus}$. However,
at the outer region, the planetesimals are continuously scattered
out of the system at $0.1$ Myr. For about $10$ Myr, there are no
survivals except at some resonances with the giant planet. As shown
in the Figure, there is a small body at the $1:2$ resonance with
Jupiter. Due to scattering amongst planetesimals, Jupiter (Saturn)
migrates inward (outward) $0.13$ AU ($1.19$ AU) toward the sun
respectively. Such kind of migration agrees with the work of
Fernandez et al. \citep{fer84} Hence, the $2:5$ mean motion
resonance is destroyed, then the ratios of periods between Jupiter
and Saturn degenerate to $1:3$. Therefore, the ratio of periods for
Jupiter, small body and Saturn is approximate to $1:2:3$. In the
$0.5 - 4$ AU, when the eccentricities of planetesimals are excited,
planetesimals are able to accrete material from wide radial
direction. The plenty of water material of the terrestrial planet in
the Habitable Zone may be transferred from the farther places by
this mechanism.

Fig. 1(b)  is illustrated for simulation 2b. In comparison with Fig.
1(a), it is apparent that planetesimals are excited more quickly at
the $3:2$ ($3.97$ AU), $2:1$ ($3.28$ AU) and $3:1$ ($2.5$ AU)
resonance location with Jupiter. The several characteristic
timescales are the same as simulation 2a for the bodies within
Jupiter. 4 planets are formed in simulation 2b, the changes of
position of Jupiter and Saturn behaves like  simulation  2a. We
point out that  simulations 2a and 2b share the initial conditions,
but the only difference in them is whether we consider the
self-gravitation among the outer planetesimals.  There is a little
gathered planetesimals survival over $400$ Myr among $7 - 8$ AU,
located in the area of $2:3$ ($6.63$ AU) and $1:2$ ($8.03$ AU)
resonances with Jupiter. The detailed results for whole simulations
that the reader may refer to \citep{Zha09}.

The production efficiency of the terrestrial planet in our model is
high, and the accretion rate inside Jupiter is $60\% - 80\%$ in the
simulations. $3 - 4$ terrestrial planets formed in $50$ Myr. 5 of 6
simulations have a terrestrial planet in the Habitable Zone ($0.8 -
1.5$ AU). The planetary systems are formed to have nearly circular
orbit and coplanarity, similar to the solar system (see Table 1). We
suppose that the above characteristics are correlated with the
initial small eccentricities and inclinations. The concentration in
Table 1 means the ratio of maximum terrestrial planet formed in the
simulation and the total terrestrial planets mass. It represents
different capability on accretion. The average value of this
parameter is similar to the solar system. Considering the
self-gravitation of planetesimals among Jupiter and Saturn, the
system has a better viscosity, so that the planetesimals will be
excited slower. The consideration of self-gravitation may not change
the formation time scale of terrestrial planets, but will affect the
initial accretion speed and the eventual accretion rate.

\begin{table}
\def~{\hphantom{0}}
  \begin{center}
   \caption{Properties of terrestrial planets
     from different systems
    }
  \label{table1}
\begin{tabular}{rcccccc}
\hline

System  &accretion rate    &$n$   &$\bar{m} (m_\oplus)$
  &concentration  &$\bar{e}$  &$\bar{i} (^\circ)$\\

\hline
   1a & 73.2518\%    & 3   & 1.8313 & 0.4606 & 0.1381 &  7.6963 \\
   1b & 80.3853\%    & 3   & 2.0096 & 0.4262 & 0.0937 &  1.7790 \\
   2a & 59.8322\%    & 3   & 1.4958 & 0.8116 & 0.2108 & 16.9117 \\
   2b & 72.9779\%    & 4   & 1.3683 & 0.4299 & 0.0999 &  5.1415 \\
   3a & 65.1098\%    & 3   & 1.6277 & 0.5337 & 0.2063 &  5.9153 \\
   3b & 66.9694\%    & 3   & 1.6742 & 0.5040 & 0.1839 &  5.2447 \\
1a-3b & 69.7544\%    & 3.2 & 1.6678 & 0.5276 & 0.1554 &  7.1148 \\
solar &  -           & 4   & 0.4943 & 0.5058 & 0.0764 &  3.0624 \\
\hline
\end{tabular}\normalsize
\end{center}
\end{table}

\section{Summary and Discussion}
We simulate the terrestrial planets formation by using two-planet
model. In the simulation, the variations of the mass of outer
planet, the initial eccentricities and inclinations of embryos and
planetesimals are also considered. The results show that, during the
terrestrial planets formation, planets can accrete material from
different regions inside Jupiter. Among $0.5 - 4.2$ AU, the
accretion rate of terrestrial planet is $60\% - 80\%$, i.e., about
$20\% - 40\%$ initial mass is removed during the progress. The
planetesimals will improve the efficiency of accretion rate for
certain initial eccentricities and inclinations, and this also makes
the newly-born terrestrial planets have lower orbital
eccentricities. It is maybe a common phenomenon in the planet
formation that the water-rich terrestrial planet is formed in the
Habitable Zone. The structure, which is similar to that of solar
system, may explain the results of disintegration of a terrestrial
planet. Most of the planetesimals among Jupiter and Saturn are
scattered out of the planetary systems, and this migration caused by
scattering \citep{fer84} or long-term orbital evolution can make
planets capture at some mean motion resonance location. Accretion
could also happen a few times between two planets if the outer
planet has a moderate mass, and the small terrestrial planet could
survive at some resonances over $10^8$ yr time scale. Structurally,
Saturn has little effect on the architecture inside Jupiter, owing
to its protection. However, obviously, a different Saturn mass could
play a vital role of the structure outer Jupiter. Jupiter and Saturn
in the solar system may form over the same period.

\begin{acknowledgments}
This work is financially supported by the National Natural Science
Foundations of China (Grants 10973044, 10833001, 10573040, 10673006,
) and the Foundation of Minor Planets of Purple Mountain
Observatory.
\end{acknowledgments}


\begin{thebibliography}{}

\bibitem[Brice\~{n}o(2001)]{bri01}
Brice\~{n}o, C. et al., 2001, \textit{Science}, 291, 93

\bibitem[Chambers(1999)]{cha99}
Chambers, J. E. 1999, \textit{MNRAS}, 304, 793

\bibitem[Chambers(2001)]{cha01}
Chambers, J. E. 2001, \textit{Icarus}, 152, 205

\bibitem[Fernandez \& Ip(1984)]{fer84}
Fernandez, J. A., \& Ip, W. H. 1984, \textit{Icarus}, 58, 109

\bibitem[Ida \& Lin(2004)]{Ida04}
Ida, S., \& Lin, D. N. C. 2004,  \textit{ApJ}, 604, 388

\bibitem[Ji et al.(2003)]{Ji03}
Ji, J. H., et al.2003,  \textit{ApJ}, 585, L139


\bibitem[Kokubo \& Ida(2002)]{kok02}
Kokubo, E., \& Ida, S. 2002,  \textit{ApJ}, 581, 666

\bibitem[Lee \& Peale(2002)]{Lee02}
Lee, M. H., \& Peale, S. J. 2002, \textit{ApJ}, 567, 596

\bibitem[Lissauer (1993)]{Lis93}
Lissauer, J.J. 1993, \textit{ARAA}, 31, 129

\bibitem[Nagasawa \& Ida(2000)]{nag00}
Nagasawa, M., \& Ida, S. 2000,\textit{AJ}, 120, 3311

\bibitem[Raymond, Quinn \& Lunine(2004)]{ray04}
Raymond, S. N., Quinn, T., \& Lunine, J. I. 2004, \textit{Icarus},
168, 1

\bibitem[Raymond, Quinn \& Lunine(2006)]{ray06}
Raymond, S. N., Quinn, T., \& Lunine, J. I. 2006, \textit{Icarus},
183, 265

\bibitem[Raymond, Mandell, \& Sigurdsson(2006)]{ray06b}
Raymond, S. N., Mandell, A. M., \& Sigurdsson, S. 2006, Science,
313, 1413

\bibitem[Safronov (1969)]{Saf69}
Safronov,V.S. 1969, Evolution of the Protoplanetary Cloud and
Formation of the Earth and the Planets (Moscow:Nauka)

\bibitem[Mayor \& Queloz(1995)]{May95}
Mayor, M.,  \& Queloz, D.  1995, \textit{Nature}, 378, 355

\bibitem[Wetherill(1990)]{Wet90}
Wetherill, G. W. 1990, \textit{Ann. Rev. Earth Planet Sci.}, 18, 205

\bibitem[\protect\citeauthoryear{Zhang \& Ji}{2009}]{Zha09}
Zhang  N.,  Ji  J.,  2009, \textit{ Science  in  China  Series G},
52(5), 794



\end{thebibliography}
\end{document}